# Tailored surfaces of perovskite oxide substrates for conducted growth of thin films

Florencio Sánchez, Carmen Ocal, and Josep Fontcuberta

Institut de Ciència de Materials de Barcelona (ICMAB-CSIC), Campus UAB, 08193 Bellaterra, Barcelona, Spain.

E-mail: fsanchez@icmab.es, cocal@icmab.es, fontcuberta@icmab.es

ABSTRACT

Oxide electronics relies on the availability of epitaxial oxide thin films. The extreme flexibility of the chemical composition of $ABO_3$ perovskites and the broad spectrum of properties they cover, inspire the creativity of scientists and place perovskites in the lead of functional materials for advanced technologies. Moreover, emerging properties are being discovered at interfaces between distinct perovskites that could not be anticipated on the basis of those of the adjacent epitaxial layers. All dreamed new prospects require the use of suitable substrates for epitaxial growth. Perovskite single crystals are the workhorses of this activity and understanding and controlling their surface properties have become critical. In this tutorial review we will chiefly focus on the impact of the morphology and composition of the surface of $ABO_3$ perovskite substrates on the growth mechanisms and properties of thin films epitaxially grown on them. As $SrTiO_3$ is the most popular substrate, we will mostly concentrate on describing the current understanding and achievements for it. Illustrative examples of other perovskite substrates ($LaAlO_3$, LSAT and $DyScO_3$) will be also included. We will show that distinct chemical terminations can exist on the surfaces used for growth and we will review methods employed either to select the most appropriate one for specific growth to allow, for instance, tailoring the ultimate outmost epilayer, or to induce self-ordering to engineer long-range nanoscale patterns of chemical terminations. We will demonstrate the capacity of this knowledge by the growth of low-dimensional organic and inorganic structures.

Key learning points
(1) What the surface of oxide perovskite single crystalline substrates for thin film growth looks like.
(2) Can we get surfaces with a single atomic termination?
(3) The choice of substrates with different orientation and chemical terminations. What we do know.
(4) Self-ordering of distinct chemical terminations of oxide substrates: an opportunity for engineered epitaxial growth.
(5) Strategies for in situ lateral arrangement of 0D and 1D nanostructures.

## 1. Introduction

Oxides are the most abundant minerals in the Earth's crust and mantle, forming minerals with different degrees of structural complexity, from the simplest rock-salt to the more complex silicates. Among them, perovskite oxides having an $ABO_3$ building block, where A and B are metallic cations, constitute a very important family with a rather simple structure but admitting an extremely large variety of cations. $CaTiO_3$ was the first discovered "perovskite" by Gustav Rose (1829) in the Ural Mountains, and named in honor of Count Lev Alexander Von Perovski (1792–1856), a Russian nobleman that combined his passion for geology with politics under Tsar Nicholas 1st. The perovskite structure is formed by a network of corner-sharing large oxygen octahedra coordinating the B metallic ions (BO6) with typical unit cells of about 0.4 nm. The center of the resulting simple structure, is occupied by the A cation which thus has a dodecahedral oxygen coordination. As recognized early by V. M. Goldschmidt (1926), this simple picture becomes much richer and more complex when one realizes that the inner cage of A cations can be much larger than the size of the ions sitting in it. Indeed, he discovered that when the cations in the A position are much smaller than the size of the cage made by the cornersharing $BO_6$ network, the perovskite structure deforms by tilting the $BO_6$ octahedra, moving the oxygen anions towards the central A cation, reducing the corresponding Madelung energy and shrinking the available volume for A cations. The overall result is that the unit cell deforms and the structure is no longer cubic but of reduced symmetry. The Goldschmidt criterion of the relative stability of perovskites and the transition from cubic to lower symmetry structures – orthorhombic, rhombohedral or hexagonal when reducing the relative size of the A cations versus B cations – has guided decades of solid state engineering of perovskite properties.

Along this fruitful history, it has become clear that the perovskite structure allows an enormously large variety of combinations of A and B cations and, not surprisingly, perovskite oxides display an unrivalled broad span of properties that have greatly impacted science and technology in the last decades. Piezoelectricity and ferroelectricity, high permittivity, magnetism and superconductivity, electrical conductivity and insulating behaviour, optical transparency or beautiful colors in gems, are properties that can be found within this broad family of compounds. Solar fuel cells, sensors and piezoactuators, magnetic or ferroelectric memories, and energy harvesting devices are examples of current applications of perovskite oxides that are deeply affecting our daily life. Oxides appear now as a new opportunity and maybe a rival for silicon electronic technology, when Moore's law and the corresponding scaling down approach their limits. Indeed, the multifunctional character of perovskite transition metal oxides may offer a new possibility for "more than Moore" in one chip. Progress in this direction requires the use of perovskite oxide thin films only a few nanometers thick. Examples[1] include the currently used ferroelectric random access memories (FeRAMs), fuel cells, magnetic tunnel junctions in magnetic memories (MRAMs) and the much investigated memristors for resistive RAMs or to mimic neuronal networks.

The progress on the growth of oxide thin films rapidly progressed beyond the late 1980's, mainly by developments in sputtering, pulsed laser deposition (PLD) and molecular beam epitaxy techniques, triggered by the rush of the high-temperature superconducting perovskite-based cuprates. This activity soon demonstrated that the perovskite thin films could be effectively grown on appropriate substrates. It was early recognized that high quality thin film epitaxies could be achieved using single crystalline perovskite substrates, which thus became the workhorse of the progress. It was also soon realized that, as much as it was already known from the semiconductor technology, the distinct unit cell parameters of substrates and growing materials could be exploited as a way to impose strain on the films to modify their properties. Thus, for instance, doubling the critical temperature of the $La_{1-x}Sr_xCuO_4$ superconductor,[2] when grown on appropriate strain-imposing substrates, constituted a hallmark evidencing the potentiality of heteroepitaxies for tailoring the properties of oxide thin films. The development of high pressure reflection high energy electron diffraction (RHEED) for in situ control of the growth of oxide thin films, which usually requires relatively high oxygen pressure, was another hallmark bringing oxide thin-film technology to its present maturity.[3]

Epitaxial oxide thin film growth requires the use of crystalline substrates whose control is challenging and appears now as the Achilles tendon of the whole growth process, dramatically affecting the thin film functionalities. Available commercial $ABO_3$ perovskite substrates are mechanically cut along a certain crystallographic direction and subsequently polished. Both, unavoidable cut/polishing imperfections and natural cleavage of the perovskite crystal, lead to a situation where multiple atomic terminations can coexist at the crystal surface severely affecting the growth of epitaxial films on top. This issue is of major relevance in the case of ultrathin transition metal oxides (TMO) oxides films displaying ferroic properties where the typical length scale for interactions is within the nanometric range, comparable to the perovskite unit cell and therefore extremely sensitive to local changes of stacking order of film layers. Last but not least, the TMO perovskite structure admits large variations in cationic composition, in their valence state and stoichiometry. This fact, which is at the root of the richness of the properties of transition metal oxide perovskites, is at the same time a crucial aspect to address when considering interfaces between films and substrates. Indeed, structural, chemical or electrostatic driven reconstructions may take place at interfaces, thus leading to emerging properties. The observation of a high mobility electron gas at the interface between two broad band gap insulators ($SrTiO_3$ and $LaAlO_3$) is probably the most celebrated example.[4]

The goal of this review is to highlight the impact that the existence of different chemical terminations on single crystalline perovskite substrates has on the properties of the ultrathin layers grown on top. With this aim, this review has been conceived to be particularly descriptive of the substrate surface-termination issues related to epitaxial oxide thin film growth and does not pretend to be either exhaustive or including other important issues for which rather vast literature exists. Thus, for instance, strain effects on oxide thin films are not reviewed here. To avoid divergence we nearly exclusively concentrate on describing the current knowledge and achievements on the control and exploitation of the surface terminations of $SrTiO_3$ (STO) as it is the substrate for which substantial knowledge has been already obtained. Still, as it will be shown here, the understanding and control of the actual STO crystal surface is incomplete, and there is much room for revision and new discoveries. In addition and to make clear the universality of some of the findings described for STO, results and properties of other increasingly demanded substrates such as $LaAlO_3$, $La_{0.18}Sr_{0.82}Al_{0.59}Ta_{0.41}O_3$ and $DyScO_3$ will be briefly described. The dependence on crystal orientation will be addressed as well and, in particular, it will be shown that in some cases surfaces with a single atomic termination (AO or $BO_2$) or self-ordered $AO/BO_2$ can be obtained. To round off following the illustrative purpose of this review, it will be shown in Sections 3 and 4 that this knowledge can be exploited to conduct the growth of subsequent layers, to control their surface-termination properties and to assist the growth of low dimensional structures.

When addressing the properties of surfaces of single crystals our regard is directed towards researchers working on oxide thin film growth, heavily concerned with the oxide/oxide interface quality in terms of well-defined chemistry and structure. The latter is the reason why thin film growers commonly use proximity probes, such as atomic force microscopy (AFM), to explore crystal surface morphology and, in a relatively fast way, to assess the surface quality prior and after growth. However, AFM goes beyond just imaging relief features and derived techniques are nowadays employed to extract relevant information and new understanding (tribology, electrical conductivity, etc.) on surfaces. Accordingly, while leaving aside other more sophisticated surface sensitive techniques, we will strengthen here the rich information on surface properties that can be extracted by relatively conventional AFM techniques, worldwide accessible.

## 2. Single crystalline oxide substrates: control of morphology and chemical termination

Lattice parameter is usually the first criterion to consider in the selection of a substrate for epitaxial growth of an oxide film. The low lattice mismatch with most perovskite oxides with interesting magnetic, electric and dielectric or superconducting properties has favored single crystalline STO to become the most popular substrate. The use of substrates with lattice parameter and/or crystal symmetry differing from that of the material to be grown makes possible tailoring lattice strain of oxide films as well as extending the number of oxides epitaxially grown. Control of the lattice strain, either by selecting appropriate substrates or suitable buffer layers, allows tailoring the properties of functional oxides, and the term ''strain engineering'' has become usual. The outstanding progress in the research of functional oxide films has been also possible thanks to the extraordinary accuracy in their epitaxial growth, permitting atomic control of the deposition process and achieving two-dimensional (2D) growth. As a result, engineered interfaces among dissimilar oxides are now within reach, thus expanding the prospects of complex oxides by including and exploiting their interfaces.[5] The 2D electron gas (2DEG) formed at the $LaAlO_3/SrTiO_3$ interface[4] and the strain-induced extremely high carrier mobility in $ZnO/Mg_xZn_{1-x}O$ heterostructures[6] are probably the most illustrative examples. The extreme sensitivity of properties of oxide films to strain, crystal symmetry, defects and interface chemistry imposes stringent requirements on their epitaxial growth. For this purpose, the substrate to be selected is no longer a passive element simply permitting epitaxy and inducing strain of the film, but other selection criteria should be considered; among them we focus here on: (i) substrate miscut, and (ii) chemical termination.

Substrate miscut is unavoidable when a single crystal is cut and polished, with the macroscopic top surface intentionally or unintentionally misoriented with respect to the closest low index (hkl) plane. A single crystal substrate surface (Fig. 1) presents terraces of average width L and separated by steps of height H (H a multiple of interplanar distances along the [hkl] direction), with the out-of-plane miscut angle (so-called: miscut angle) a = atan(H/L). Steps can be aligned forming an azimuth in-plane angle f (the in-plane miscut angle, also called miscut direction) with respect to the closest crystal direction with lower step energy. Substrate steps are preferential sites for adsorption of atoms during epitaxy, and thus they play a key role to tune 2D film growth (layer-by-layer or step flow) and their persistence against 3D growth instabilities. Though miscut angles a usually range from 0.051 to 0.31, corresponding to terrace widths in the 70–500 nm range, substrates with lower or higher a are available; these surfaces are called singular and vicinal, respectively (see sketches in Fig. 2a–c). Atoms at steps have lower coordination than atoms at terraces, increasing the total surface energy. A lower energy state can be achieved by step bunching, resulting in wider terraces separated by steps n x H in height (Fig. 2d). Step bunching can be favored by annealing vicinal crystals, or by high temperature deposition of a film on a vicinal substrate. On the other hand, the in-plane orientation of the steps, i.e. the miscut direction f, can be more or less close to that of a low energy step direction. If f is small, steps will be parallel, even up to micrometric distances, as sketched in Fig. 2e.[7] However, if f is large, steps can zig-zag greatly, with alternating kinks and straight segments reducing locally the width of terraces (Fig. 2f).

The relevance of the surface chemical termination can be illustrated considering (001) surfaces of perovskites, which can be terminated in either the AO or $BO_2$ atomic planes. A substrate is single terminated when all terraces are fully terminated in either AO or $BO_2$ atomic planes (Fig. 3a and b). However both terminations can coexist, their balance depending on their relative surface energy. Random distribution of the minority termination is expected in a polished crystal (Fig. 3c), but these regions may coalesce by annealing to reduce step energy (see sketch in Fig. 3d). It will be shown below that the control of the chemical termination of a substrate permits either selecting a particular atomic stacking at an interface in the case of single termination, or driving a self-organized growth of nanostructures in the case of mixed termination.

## 2.1 $SrTiO_3$(001): single chemical termination

STO, with a lattice parameter of 3.905 A, is formed by stacking SrO and $TiO_2$ atomic planes along the [001] direction. In as-received polished STO(001) single crystalline substrates, both SrO and $TiO_2$ terminations coexist. Commonly, SrO termination is the minority, being about 5–25% of the total surface area.[8] The STO(001) surface has been widely investigated, and multiple reconstructions, nanostructures, cationic off-stoichiometry and precipitates at the surface have been observed depending on the chemical and/or thermal treatments (see for instance ref. [9]). The large variety of reported results signals the strong sensitivity of STO(001) surfaces on the processing conditions, including interaction with chemicals, annealing (gas, temperature, time and rate) and the microstructure (defects, dopants, impurities, Sr offstoichiometry) of the single crystal.

**2.1.a Processing for single $TiO_2$-termination.** The extended use of STO(001) as substrate motivated efforts to obtain single chemical termination, and pure $TiO_2$-termination (denoted here TiO2–STO) was obtained in 1994, by using selective chemical etching.[8] A buffered NH4F–HF (BHF) solution was used to dissolve selectively the SrO-termination. It was found that the accurate control of the pH of the solution was critical; indeed, pH 4 5 resulted in inefficient SrO etching, whereas pH o 4 caused dislocation etching pits. Moreover, it was found that the etching time to completely remove the SrO layer, depends on specific characteristics of the STO(001) substrate (percentage of SrO termination, miscut angle) and that etching may also cause quick formation of extended defects.[10] It was subsequently found that sinking the substrate in water to form $Sr(OH)_2$, prior to the BHF etching, made easier the elimination of the SrO termination.[11] Therefore, the etching time and/or HF concentration can be reduced, resulting in less damage in the crystal. Finally the etched crystals were annealed at 950 °C for about 1 hour to obtain well-defined topography of terraces and parallel steps. The method is used today as the standard recipe to obtain atomically flat terraces. It has been recently proposed that single $TiO_2$-termination can also be obtained after the simple process of boiling in water to form Sr-hydroxide followed by annealing in air (950 °C, 8 hours),[12] or by a first annealing (air, 1000 °C, 1 hour), followed by ultrasonic immersion in water, and a second annealing under the same conditions as the first one.[13] If such treatments were effective they would allow reducing defects in treated crystals. Nevertheless, Ohnishi et al. claimed that there is an intrinsic instability of BHF etched STO surfaces when the crystals are heated.[14] The authors used coaxial collision ion scattering spectroscopy to measure the composition of the terminating layer of etched crystals as a function of annealing temperature and detected significant Sr content in samples heated above 300 °C. However, AFM topographic images did not evidence minority SrO-terminated regions even in samples annealed at temperatures above 1000 °C, in contrast with usual morphologies of annealed STO(001) crystals having both terminations (described below in Section 2.2). In a more recent study, Herger et al.[15] used in situ surface X-ray diffraction to determine the surface structure of TiO2 terminated STO at $10^{-3}$ Pa of oxygen and a typical perovskite thin film growth temperature (750 °C). They found a Ti-rich (1x1) surface, terminated with two $TiO_2$ atomic layers, and with atomic displacements down to three unit cells.[15]

**2.1.b Processing for single SrO-termination.** Enrichment of Sr at the surface of STO by high temperature annealing permits obtaining single SrO-terminated STO(001) substrates (denoted here SrO–STO). The effects of the Sr diffusion towards the STO(001) surface depend on the annealing conditions. Szot et al.[9] annealed crystals (200 Torr of oxygen, 900–1000 °C, 24 h) and reported that Ruddlesden–Popper phases were formed at the crystal surface; further increasing the temperature and annealing time, SrO microcrystals developed. It was recently shown[16] that simple thermal treatment in air above ~1200 °C causes progressive enrichment of Sr while preserving the surface low roughness. The progressive transformation from $TiO_2$-terminated to SrO-terminated STO(001) surfaces by thermal/time annealing is illustrated in Fig. 4. Fig. 4a shows AFM topographic and phase images of STO(001) crystals, first etched chemically with BHF to obtain single $TiO_2$-termination and then annealed in air at 1300 °C for 2 hours. After annealing for 2 h the surface presents atomically flat terraces and steps ~0.39 nm high, one unit cell (u.c.) of STO, signaling single $TiO_2$-termination within the AFM resolution. In agreement, the phase-lag image does not show contrast between terraces. However, longer annealing time (12 h) leads to evident changes in the surface (Fig. 4b). The observation of an array of narrow terraces edging with low-lying leaf-shaped regions separated by steps of ~0.5 u.c. height indicates the appearance of the SrO chemical termination. Indeed, a strong contrast is observed in the corresponding phase-lag image between these regions confirming that these two kinds of terraces have distinct chemical terminations. This corroborates the SrO surface enrichment by high temperature annealing; remarkably, when using the described conditions, the formation of SrO microcrystals is avoided. Interestingly, longer annealing (72 h, Fig. 4c), leads to a flat surface without visible lag phase-contrast, indicating a single termination. The contrast observed at the step edges is due to the sharp height differences and is not caused by a chemical effect. Therefore, within the resolution of the standard AFM microscopes, the resulting surface is SrO single-terminated. Furthermore, the corresponding topographic image shows atomically flat terraces with step edges facetted along the [100] and [110] directions and step heights of integers of the STO u.c. (either 1 or 2 u.c.), suggesting the conservation of the perovskite STO structure. These SrO–STO(001) surfaces permit the epitaxy of complex oxides by layer-by-layer growth, as described in Section 3.

## 2.2 SrTiO$_3$(001): two chemical terminations patterns

In as-received STO(001) crystals, the minority SrO termination cannot be resolved with standard AFM, indicating a small lateral size. However, its presence is revealed by the otherwise unexpected substantial surface roughness on the terraces of untreated samples.[17] The small lateral sizes of SrO regions imply long step edges separating $TiO_2$ and SrO terminated regions and thus coalescence of the minority SrO-terminated regions could occur at high temperature to reduce the overall surface energy. The final result would be a surface with separated SrO and $TiO_2$ regions. This is recurrently observed after annealing at ~1000 °C,[11,17,18] a temperature below the onset of massive diffusion of Sr from the bulk. In a recent study, a STO(001) crystal was annealed at a moderate temperature (1100 °C) over a short time (30 min). The chemical nanopatterning by surface diffusion is not complete and simultaneous topographic and lateral force monitoring permits visualizing the surface evolution.[18] Fig. 5a shows the incipient state of the self-assembly of the chemical terminations. The stepped surface with flat $TiO_2$-terminated terraces separated by ~0.4 nm high steps (one u.c. in height) present meandering step-edges and some rounded holes in the flat terraces (Fig. 5a and b). The holes, with diameters from tens of nm to a few hundred nm, are located in the terraces at diverse distances from the descending step edges; of relevance is that all of them are about 1.5 u.c. (~0.6 nm) deep. Since SrO termination is a minority part of the surface of untreated STO(001) crystals,[8] these 1.5 u.c. deep holes should be vacancy-islands with SrO termination.

Holes of different sizes are the result of coalescence of preexisting SrO terminations at the surface of the as-received substrates. The dynamics of the holes' diffusion can be better appreciated by comparing their sizes and positions. In Fig. 5a and b some holes have been labeled (from 1 to 8) at progressively shorter distances from the step edges. Thus, several stages of the vacancy islands' diffusion can be seen as frozen frames of a time-resolved surface evolution. Once a specific vacancy island is close enough to the step edge, the resulting neck is subsequently narrowed (3, 4) until it is broken to minimize the island perimeter plus step border (5, 6). At this stage, the total ledge relaxation gives rise to leaf-shaped structures (7, 8). Substrates annealed at the same temperature but with longer time (Fig. 5c) do not show larger holes at the terraces but the SrO-terminated regions form elongated structures along the steps and have 0.5 u.c. and 1.5 u.c. depth with respect to the lower and upper terrace levels, respectively. It can be appreciated that the SrO structures are isolated in some samples, whereas in others are connected along each step, forming long one-dimensional (1D) narrow nanoribbons (Fig. 5c). We end by remarking that coalescence of SrO termination by annealing as-received STO(001) substrates signals that the chemical termination striped pattern lowers the surface energy of the crystal.

## 2.3 SrTiO$_3$(110) and (111), and other oxide substrates

The popularity of STO substrate has been focused on STO(001) oriented crystals, with less interest in other surfaces. Today, epitaxy on substrates with other orientations is perceived as an opportunity to extend the range of properties of the highly anisotropic oxides.[19] However, the fabrication of high quality epitaxial oxide films having other orientations is in general challenging due to the fact that in perovskites, usually the (001) surfaces have the lowest energy. Moreover, the (110) and (111) surfaces of STO are polar, usually presenting reconstructions or covered by adsorbates, thus challenging the understanding and control of their surfaces. Ideal STO(110) can be terminated in $SrTiO^{4+}$ or $O_2^{4-}$ layers, and STO(111) in $SrO_3^{4-}$ or $Ti^{4+}$ layers. The atomic structure of these surfaces can be complex and highly dependent on processing conditions, but both STO(110) and STO(111) crystals can be made to present flat surfaces after appropriate treatment.

**2.3.a SrTiO3(110).** Mukunoki et al.[20] used a two-stage annealing process to obtain single-terminated atomically-flat STO(110) substrates. The first anneal was performed at high temperature (1000 °C) and low oxygen pressure (5 x $10^{-5}$ to 5 x $10^{-7}$ Torr) to induce oxygen vacancies near the surface to compensate for the otherwise charged surface. The surface neutrality and the high temperature permitted achieving an atomically flat surface. Then, a second anneal stage (550 °C, 1 x $10^{-4}$ Torr) permitted recovery of the oxygen contents in the crystal bulk while preserving the surface quality.

Later on, Herranz et al.[21] showed that a single annealing in air at 1000–1100 °C produces similarly flat surfaces (AFM topographic images are shown in Fig. 6a and b). Although steps are not straight, the terraces are very flat and step heights (Fig. 6d) signal single termination. The RHEED pattern in Fig. 6c shows Bragg spots along the 0th and the 1st Laue circles and Kikuchi lines, confirming high structural quality over a large area. On the other hand, scanning tunneling microscopy of Nb-doped STO(110) annealed in ultrahigh vacuum for 2 hours at 875 °C, 1100 1°C or 1275 °C showed 3 x 1, 4 x 1 and 6 x 1 reconstructions, respectively; in all cases without {100} microfacets and with steps about ~0.27 nm high (the nominal height between terraces for a single termination).[22] Biswas et al.[23] reported that BHF etching and annealing for 3 hours at 1000 1C under oxygen flow allowed to obtain STO(110) crystals presenting, within the AFM resolution, flat terraces. These authors used mass spectroscopy of recoiled ions to quantify differences in the topmost atomic layers and found an increased Ti/Sr ratio in the treated sample with respect to an as-received crystal, indicating selective etching of Sr.

**2.3.b $SrTiO_3$(111).** STO(111) crystals with flat terraces have been obtained either by annealing in air (1100 °C, 2 hours)[21] or vacuum (850 °C, 30 min).[24] The RHEED patterns and the AFM images of samples annealed in air (Fig. 6) confirm a high quality surface over large areas, with flat terraces and parallel steps of uniform height of ~0.25 nm, thus signalling single termination.[21] Similarly, scanning tunnelling microscopy experiments on Nb doped crystals annealed in vacuum showed only steps ~0.23 nm high, signaling single chemical terminations.[24] The authors observed two reconstructions and argued that surface reconstruction involved oxygen vacancies to neutralize the nominally polar STO(111) surface. Several groups have also treated STO(111) crystals by BHF etching and thermal annealing (1000–1050 °C, oxygen atmosphere).[23,25] The morphology was similar to that obtained after thermal annealing without etching.[21]

**2.3.c Other oxide substrates.** The use of oxide substrates alternative to STO is increasing quickly and thus the control of their surface is necessary. Here we briefly comment on some of the most usual substrates beyond STO(001): (001)-oriented $La_{0.18}Sr_{0.82}Al_{0.59}Ta_{0.41}O_3$ (LSAT), $LaAlO_3$ (LAO), and $DyScO_3$ (DSO), the two latter are indexed here as pseudocubic. All present ABO3 perovskite-like crystal structures with alternating AO and $BO_2$ charged planes along [001] and lattice parameters of 3.87 Å, 3.79 Å, and 3.94 Å, respectively. LAO(001) is the most studied, and several groups (see for instance ref. [26]) have concluded that simple thermal annealing results in single termination. However other studies have detected minority termination and there is also controversy identifying the main (single or majority) termination as $AlO_2$ or LaO.[27] Single termination by annealing was also claimed for LSAT(001),[28] but it was recently shown that after annealing in a wide range of conditions (time and temperature), the crystal surfaces present both terminations,[29] forming striped patterns similar to those observed in annealed STO(001). A striped pattern of AO and $BO_2$ terminations was also reported for annealed DSO,[30] and it has been found that etching with a solution of NaOH and distilled water allows obtaining single termination.[30]

# 3. Chemical termination control at surfaces and interfaces of oxide films

The use of single terminated oxides substrates is a prerequisite to obtain the highest quality oxide thin films. Indeed, using only $TiO_2$–STO substrates it was possible to obtain perfect layer-by- layer growth of homoepitaxial SrTiO3 films.[8,11] The extreme sensitivity of thin film growth on substrate termination is well illustrated by the growth by PLD of SrRuO$_3$ on TiO$_2$–STO. During the first stages of growth, the TiO$_2$ termination is fully transformed into SrO due to a surface Sr enrichment associated with the high volatility of Ru$_x$O$_y$. As a consequence of this conversion the subsequent SrRuO$_3$ grows, in stacking and stoichiometry, as when growing SrRuO$_3$ on SrO–STO.[31] In other words, the chemical termination of the substrate, SrO–STO or TiO$_2$–STO, does not impose a different chemical termination at the surface of the SrRuO$_3$ film. We remark that these observations are related to the volatility of Ru oxides and thus may not occur in other systems. Indeed, it will be shown in the following that, in general, the chemical terminations of the substrate determine the chemical terminations of the growing layer providing a unique tool to obtain in-plane control of the chemical and thus electronic properties of epitaxial oxide thin films.

### 3.1 Control of chemical termination in oxide films

Manganite perovskite thin films, i.e. La$_{2/3}$Sr$_{1/3}$MnO$_3$ (LSMO) and related compounds have been much investigated due to the unrivalled, rich phase-diagram of electric and magnetic properties. When growing LSMO on STO or other substrates, such as LSAT, it is of relevance to determine whether the substrate terminations replicate at the film free surface or not. This question can be recast by asking if the AO or BO$_2$ terminations of the films mimic those of the substrate. The answer to this question could be extremely relevant when integrating LSMO-like thin films in sandwich structures, such as tunnel junctions, where properties are known to be largely determined by the chemical nature of the interface.

A first insight into this crucial issue can be obtained by growing LSMO films on STO or LSAT substrates, assessing and controlling the growth mechanism by RHEED and subsequently exploring the morphology and the chemical and electric properties of the resulting films.[32] In Fig. 7a we show an AFM topographic image of a thermally treated LSAT(001) substrate where terraces and steps differing in altitude by 0.5 u.c., can be observed (Fig. 7b); this spacing indicates the presence of the two different terminations (AO and BO$_2$). This is better seen in the lateral force map (inset in Fig. 7a) where the contrast between terraces separated by steps 0.5 u.c. high confirms their different chemical nature. Fig. 7d shows the topographic AFM image of a 4 monolayer (ML) thick LSMO film grown by PLD. The AFM image shows that the film replicates the morphology of terraces and steps of the underlying substrate as expected from the layer-by-layer growth inferred from the oscillations of the RHEED intensity.[32] Correspondingly, the line scans of the topographic profile (Fig. 7e) display only 0.5 and 1.5 u.c. high steps. The contrast in the lateral force map (inset in Fig. 7d) confirms that the two sets of terraces correspond to different chemical terminations. The sketches in Fig. 7c and f illustrate the surface self-structuration of the LSAT substrate and that of a LSMO film grown on it, respectively.

Even more interesting, the Kelvin probe microscopy images of the LSMO free surface evidence a modulated contrast (Fig. 8b) that mimics the friction and topography images (Fig. 8a), thus signaling a different local work function for each LSMO termination. These results, and similar ones obtained for LSMO grown on STO(001)[32] indicate that the free surface of manganite thin films replicates the pattern of the substrates, i.e. no termination conversion occurs, in contrast to the SrRuO$_3$ on STO(001) case discussed above and explained by the high vapor pressure of Ru$_x$O$_y$.

In short, it has been shown that when LSMO is grown on AO or BO$_2$ terminations of STO(001) and LSAT(001) substrates, the free surface of LSMO displays distinguishable LaSrO and MnO$_2$ terminations. Similarly, it has been reported that fine control of the growth process allows the replication of the substrate morphology in multilayered heterostructures as demonstrated in the case of BaTiO$_3$/LSMO/STO(001).[32] It follows that using the two different terminations of the substrates, the free surface of LSMO and BaTiO$_3$ films, and the atomic stackings (BaO/MnO$_2$ or TiO$_2$/LaSrO) at the interface, can be tailored *ad hoc*.

A beautiful illustration of this statement is obtained measuring the electronic occupancy at $x^2 - y^2$ and $z^2$ orbitals of the 3d-Mn states of the outermost MnO$_2$ layer in LSMO thin films. The X-ray absorption and L2 and L3 edges of Mn are sensitive to the symmetry of the available states. In LSMO, the relevant 3d orbitals are the $x^2 - y^2$ and $z^2$ and the X-ray absorption will be different depending on the relative energy position and electronic occupancy of these orbitals, thus giving rise to X-ray absorption linear dichroism (XLD). In thin films, two different contributions come into play determining the relative occupancy $x^2 - y^2$ vs. $z^2$ and therefore the XLD. The first one is the strain acting on the films due to the mismatch with the substrates. A tensile (compressive) strain lowers the $x^2 - y^2$ ($z^2$) orbital with respect to the $z^2$ ($x^2 - y^2$) orbital. The second contribution comes from a genuine symmetry breaking at the free surface of films that suppresses the degeneracy of $x^2 - y^2$ and $z^2$ electronic state of Mn$^{m+}$ ions, by pushing up in energy the in-plane $x^2 - y^2$ states and leveling down the out-of-plane $z^2$ orbitals. This effect arises due to the fact that the oxygen coordination polyhedron of Mn$^{m+}$ ions of the MnO$_2$ surface layer lacks one apical oxygen. This contribution would be absent if the LSMO film was terminated by a LaSrO layer, because in this case the Mn$^{m+}$ ions of the topmost MnO$_2$ will have the octahedral oxygen coordination complete.

In Fig. 9 we show the XLD data of a 4 u.c. thick LSMO film grown in layer-by-layer mode, on TiO$_2$–STO and SrO–STO substrates.[33] The dichroic peak around 655 eV of the LSMO film on TiO2–STO is largely positive, indicating a preferential occupation of eg-3d Mn$^{m+}$ electrons at 3d-$z^2$ states. This is in agreement with the fact that, on the TiO$_2$–STO substrate, the topmost layer of the LSMO film should be MnO2 terminated; the absence of one apical oxygen around Mn$^{m+}$ ions pushes the $z^2$ orbitals down in energy as described above, thus favoring its electron occupancy. The LSMO film grown on SrO–STO substrates displays a much reduced XLD around 655 eV, signaling a reduced occupancy of $z^2$ orbitals. Detailed analysis of data shows that a negative contribution, indicating preferential $x^2 - y^2$ occupancy, to the XLD should be expected due to the unavoidable tensile strain induced by the STO substrate on the LSMO.[33]

Next, we present an example of lateral modulation of the atomic stacking at the interface between LAO and STO, which allows obtaining lateral modulation of the functional properties inherent to the LAO/STO interface. When growing epitaxial LAO layers on TiO$_2$–STO substrates, above a critical LAO thickness of about 4 u.c., a highly conducting 2DEG is formed, being confined in a narrow region, a few nanometers thick, parallel to the interface.[34] The ability to self-pattern STO substrates in distinguishable TiO$_2$ and SrO terminations, only a few tens of nanometers in lateral size as described before, opens the possibility to obtain laterally confined 2DEG regions. In Fig. 10a we show an electron back-scattering image, obtained using field-emission scanning microscopy, from an LAO/STO sample prepared in such a way that 6 u.c. of LAO were grown, by RHEED-assisted PLD, on a STO substrate containing SrO and TiO$_2$ terminations. The films display the characteristic metallic conductivity at low temperature, signaling the formation of a 2DEG at LAO/TiO$_2$-interfaces.[35] Moreover, electrical resistivity measurements display large in-plane anisotropy.[35] The bright/dark contrast in Fig. 10a reflects the pattern of the underlying LAO/STO interfaces with distinct conductance at LAO-covered SrO-terminated regions (dark) and LAOcovered TiO$_2$-terminated regions (bright); interestingly, aligned along step edges. This observation is in full agreement with the formation of trenches of SrO close to step edges as described above. On the other hand, the presence of the insulating regions, aligned along step edges is at the origin of the observed anisotropic in-plane resistivity.

The AFM phase image (Fig. 10b) nicely reflects the coexistence of different terminations at the LAO film and the concomitant change of surface properties.

Before closing we would like to add that distinct growth on the chemical terminations of single crystalline substrates, as illustrated here for STO(001) and LSAT(001), may be only the tip of the iceberg of more complex and rich phenomena. For instance, it is known that different surface reconstructions, strongly depending on the initial conditions of the surface and the processing parameters (oxygen pressure, temperature and annealing time), can take place at the surface of STO(001) single crystals; some of these reconstructions, implying stoichiometry variations, also constitute templates for film growth. It has been shown, for instance, that homoepitaxial $SrTiO_3$ or $SrVO_3$ layers grow distinctly on STO(001) depending on the precise local reconstruction.[36]

## 4. Selective growth and low-dimensional nanostructures

In previous sections we have shown that the $AO/BO_2$ terminations of single crystalline $ABO_3$ substrates can self-order and, by coalescence along the step edges, they may form long ribbons of AO terminations adjacent to steps. We have shown that AO and $BO_2$ terminations allow distinctive growth of oxides replicating the morphology and the chemical terminations at their free surface. Below we will show that some oxides and organic materials deposited on self-ordered $ABO_3$ substrates grow selectively on one single chemical termination. This permits fabricating oxide nanoribbons or nanostructured organic monolayers. Next we will describe some strategies developed to achieve in situ spatial and/or orientational ordering of oxide structures with lower dimensionality, i.e. 1D nanowires and 0D dots.

### 4.1 Selective growth of oxide nanoribbons

An unexpected transition from 3D to 2D growth mechanisms with thickness was observed in films of the ferromagnetic metallic oxide $SrRuO_3$ on untreated STO(001) substrates.[37,38] It was noted[37] that $SrRuO_3$ nucleated at the step sites resulting in ''finger'' shaped islands scattered along the steps. The early coalescence of these islands along the step direction results in quasi 1D nanostructures,[38] and the critical role of the substrate steps on the 1D nanostructures permitted controlling its size by using substrates of specific miscut angle.[7] 1D $SrRuO_3$ nanostructures were also obtained on phototreated STO(001) substrates, and Ru re-evaporation at step sites with remnant SrO termination was proposed as mechanism for the $SrRuO_3$ growth mode.[39] This was confirmed[17] by depositing ultrathin $SrRuO_3$ films on STO(001) substrates treated thermally to develop the pattern of SrO and $TiO_2$ terminations described in Section 2. The long-range order and homogeneity of the SrO-terminated regions running along step-edges of these substrates can be appreciated in Fig. 11a. After a coverage of 7 ML of $SrRuO_3$ (B3 nm), the resulting film morphology (Fig. 11b) clearly mimics that of the substrate pattern, which acts as a template for replication. The quality of the long-range ordering of 1D $SrRuO_3$ nanostructures can be better appreciated in the 5 x 5 $mm^2$ image of Fig. 11c. The deep trenches formed in the $SrRuO_3$ film corresponds to uncovered SrO–STO regions which thus exhibit an insulating character. This can be confirmed by current sensing AFM measurements using a conducting tip. In Fig. 11d we show the map of current flowing between the tip and the sample, recorded by using the experimental set-up sketched in the inset of Fig. 11d using a bias voltage of V = 500 mV. The insulating nature of the self-ordered trenches sharply contrasts with the conducting nature of the $SrRuO_3$ broad stripes covering the terraces. A similar selfordered array of $SrRuO_3$ ribbons has been also recently observed when SrRuO$_3$ is grown on DyScO$_3$,[30] and similarly attributed to the selective growth of SrRuO$_3$ on the ScO$_2$ terminations.

## 4.2 Selective growth of organic layers

The possibility of using these chemically self-patterned oxide substrates for selective growth of other materials, organics in particular, is addressed here by demonstrating that they can act as platforms for nanostructuring organic-based monolayers of stearic acid (n-C17H35COOH). By choosing the appropriate solution concentration and by using a room-temperature drop-casting method, a completely selective adsorption of stearic acid at TiO$_2$ terminations can be obtained.[18] Self-assembled monolayers were observed as inferred from data in Fig. 12. The wide terraces, corresponding to TiO$_2$ terminations are covered by stearic acid whereas the darker regions, oval shaped, are uncovered. The line profile in Fig. 12 (inset), crossing two surface steps and a bare SrO terminated region, allows measuring an organic layer thickness of 1.3 +/- 0.2 nm indicating that the molecules (∼ 2 nm high) stand up tilted by 50° with respect to the surface normal. Friction force and work function measurements confirmed the distinct chemical nature of covered and uncovered regions and illustrated that the morphological, chemical, electronic and tribological properties of the STO surfaces can be locally controlled.

## 4.3 1D oxide nanowires

2D structures form when a film grows layer-by-layer or by step flow, and alternating deposition of two oxides can be used to get periodicity along the out-of-plane direction (superlattices). Vertically oriented nanowires are usually fabricated using metal catalysts, or can emerge naturally during three-dimensional growth of crystalline oxides with strong crystalline anisotropy. These types of nanowires are widely described in the literature; in this review we will focus on oxide nanowires that grow confined within the plane of a substrate ("horizontal nanowires"). Spatial order of horizontal nanoribbons can be achieved when an oxide deposited on a substrate, with a striped pattern of chemical terminations, grows selectively on one of them, as described above. Usually, nucleation occurs randomly, and thus horizontal nanowires grow with lack of long-range spatial order. However random nucleation is not the case in two-dimensional growth by step flow mechanism. Yoshimoto et al.[40] used Al$^2$O$_3$(0001) substrates with narrow terraces to confine the growth of (Mn,Zn)Fe$_2$O$_4$ and Fe$_3$O$_4$. They obtained parallel nanostructures, 15–20 nm wide and 0.5 nm high, placed along substrate steps. There is remarkable long-range order, but the two-dimensional growth mechanism limits the nanostructure shape to a height below 1 nm. Horizontal nanowires can also form when the deposited oxide presents strong in-plane anisotropy. For example, when Gd-doped CeO$_2$ (CGO) grows (011)-oriented on LAO(001) substrates, the anisotropic strain favors growth along the CGO[0-11] in-plane direction.[41] Two sets of orthogonal nanowires form due to the two-fold symmetry of the CGO(011) planes and the four-fold symmetry of the LAO(001) surface, forming labyrinth structures as they nucleated at random positions. A single orientation of CGO nanowires was induced by nanoindenting the LAO(001) substrates, with the nanowires oriented parallel to the scratched lines.[41] Alternatively, single in-plane orientation of horizontal nanowires can be achieved on substrate surfaces with two-fold symmetry, as demonstrated with In$_2$O$_3$ nanowires on YSZ(110) substrates, with perfect single orientational order although without presenting spatial order.[42]

## 4.4 0D oxide nanostructures

0D epitaxial oxide nanostructures form naturally during thin film deposition when the growth mechanism is Volmer–Weber or Stranski–Krastanow. Also, instabilities during layer-by-layer growth cause multilayered islands (usually called mounds). Islands formed by Stranski–Krastanow and Volmer–Weber mechanisms can display narrow size distribution and well defined shape with orientational order. For example, epitaxial spinel oxides on (001)-oriented cubic substrates tend to form {111}-facetted pyramidal islands because of the low surface energy of these planes.[43] These islands, due to their {111}- facetted shape, display perfect orientational ordering along [110] in-plane directions. However, since nucleation is generally a stochastic process, 0D nanostructures are randomly distributed on the surface. The strategies to achieve lateral ordering during deposition can be divided in two groups (Fig. 13): (A) ordering by engineered growth, and (B) ordering by selforganization. The first group includes two strategies, A1 and A2. A1 is based on acting artificially on the material that is being deposited, for example by means of stencils, to confine growth at specific regions. The methods based in stencils have progressed notably and patterns of oxide nanostructures with submicronic lateral size can be fabricated even at the high temperature usually required for oxide epitaxy.[44] The second strategy, A2, consists in the deposition on previously patterned substrates, aiming for selective growth at the patterned regions. The second group includes two strategies, B1 and B2. In B1, selective growth is achieved using single crystalline substrates presenting a periodic pattern of chemical terminations (see Section 2) or epitaxial films showing the cross-hatch morphology that can emerge due to the periodicity of misfit dislocations. B2 considers direct spontaneous ordering during heteroepitaxy. The driving forces include long-range elastic interactions between islands and anisotropic adatom diffusion due to the Ehrlich–Schwoebel barrier at substrate steps. These strategies, widely employed to fabricate semiconducting nanostructures, are much less used to fabricate oxide nanostructures and the underlying mechanisms are far from being fully understood. Here we focus on 0D oxide nanostructures obtained on pre-patterned substrates (A2) or formed exclusively by self-organization (B1 and B2).

Substrate steps are critical in the epitaxial growth of a thin film. In two-dimensional growth the transition from layer-by-layer to step flow mechanism is favored by increasing the miscut angle, which determines the step density and terrace width. In three-dimensional growth, steps can also favor islands nucleation, as observed for $Bi_2O_3$ on STO(001)[45] or YSZ on $Al_2O_3$(0001).[46] The relevance of substrate steps on growth depends on their orientation and height. Considering an ideally single-terminated surface, steps are usually active sites for incorporation of adatoms diffusing from the lower terrace. But, on the other hand, the Ehrlich–Schwoebel barrier introduces anisotropy in the adatoms flux, reducing diffusion from upper to lower terraces. In this case, the higher supersaturation close to descending steps increases the probability of islands nucleation close to steps of upper terraces. Oxide nanostructures can also nucleate far from the steps, as reported for $CeO_2$(001) epitaxial islands grown by chemical solution deposition on LAO(001).[47] $CeO_2$ dots did not nucleated on the steps and were exclusively found on terraces, being argued that the high interfacial energy between $CeO_2$ dots and LAO steps is the driving force for the observed spatial confinement of $CeO_2$ dots.

Pre-patterned substrates for the selective growth of semiconductors has been used to fabricate spatially ordered semiconductor nanostructures. Although in the case of oxides, this approach has not received comparable attention, it has been already shown that it may also be used for efficient control of the nucleation sites. For instance, a 2D square array of depressions, a few hundreds of nanometers wide each, can be fabricated by irradiating a STO(001) crystal with a focused ion beam; and subsequently, $Cu_2O$ was deposited on this surface.[48] It turned out that $Cu_2O$ grows three-dimensionally on STO(001), but on the patterned STO(001) the islands (~100 nm wide) nucleated exclusively at the depressions' edges, replicating the square array pattern.[48]

Recently, metallic Fe has been used as a seed to confine the growth of antiferromagnetic $LaSrFeO_4$.[49] An array of Fe dots (~200 nm in width and spacing) was fabricated on STO(001) by nanoimprint lithography and crystallized thermally. Then, by PLD of $La_{0.5}Sr_{0.5}FeO_3$, an epitaxial nanocomposite was formed spontaneously, nucleating selectively Fe on the Fe dots and $LaSrFeO_4$ on the rest of the substrate. Minimization of interface energy (the lowest corresponding to Fe on Fe, and $LaSrFeO_4$ on STO) and strain energy was proposed as the driving force for the selective nucleation. On the other hand, it has been proposed that interface energy minimization is the driving force for the selective growth of $SrRuO_3$ nanodot arrays on thermally treated LSAT(001).[29] Annealing of LSAT(001) causes self-separation of AO and $BO_2$ terminations forming a striped pattern (Section 2). $SrRuO_3$ grows epitaxially and three-dimensionally (the strong mismatch of ~1.5% between SRO and LSAT hinders 2D growth) on these patterned substrates, but nucleates exclusively on one of the terminations to form arrays of dots (Fig. 14). This result demonstrated that spatial confinement of 0D oxide nanoobjects can be fabricated entirely by self-organization processes, without lithographic stages. On the other hand, heteroepitaxial films can relax plastically forming periodic patterns of misfit dislocations (the usually called cross-hatch morphology) that could be used as natural templates. This approach has permitted obtaining square patterns of Ge quantum dots using SiGe/Si(001) samples showing cross-hatch morphology as substrates. Some perovskite films as $SrRuO_3$ (ref. [50]) develop cross-hatch morphology after proper annealing, and thus they are promising as templates for growth of oxide 0D arrays.

## 5. Conclusions and outlook

Technological progress on integrating metal oxides in nanometric thin films requires an exhaustive control of the thin film growth which largely depends on the substrate preparation and stability, as well as on the growth conditions. Advanced RHEED has largely improved the growth control, making it possible to follow and verify layer-by-layer growth of oxides, dramatically approaching the ultimate limits of quality control of semiconductor technologies. The knowledge of the substrate structure at the atomic scale is a requisite that has progressed enormously in recent years and, for instance, recipes to obtain reproducible STO(001) surfaces with controlled terminations are now rather well settled. However, the situation for other perovskite substrates and other crystal orientations is not as mature yet and more efforts are definitely required to get a full understanding of their surface chemistry, crucial for obtaining single terminated surfaces in a controlled way. This is especially true, and probably an urgent need, for $ABO_3$(111) crystal orientations, where the particular symmetry of the surface could allow investigation of so-far unexplored avenues in oxides.

The existence of single atomic termination, as required by most advanced applications and the search of emerging properties is commonly addressed by using ex situ AFM performed at room temperature. This approach is fully valid to determine the properties of the final surface but also has inherent drawbacks in reference to growth characterization. On the one hand, AFM provides local information, only partially circumvented by statistical measurements; other surface sensitive techniques with larger lateral sampling dimensions might be more appropriate for average surface homogeneity determination. For this purpose, electron diffraction (RHEED and low energy electron diffraction), surface X-ray diffraction, and X-ray photoelectron spectroscopy, among other techniques, are available. On the other hand, standard

AFM lacks enough lateral resolution to discriminate nanometric patches of different terminations. Moreover, the surface morphology is usually obtained at conditions largely differing from those of actual growth, typically performed at high temperature. In situ high temperature proximity probes and other spectroscopic techniques may constitute the next breakthrough in the field. These new tools may help monitoring in situ, during thin film growth, the chemical composition and atomic inter-diffusion, which may have a tremendous impact on the properties of ultrathin oxide layers and largely modify the properties of interfaces.

Through the present review we have shown that the presence of multiple terminations (AO or $BO_2$) can be viewed as a new opportunity of surface engineering for materials design, by exploiting the inherent tendency (driven by interface energy minimization) to merge together forming long range self-ordered structures. Due to the distinct chemical reactivity of these terminations, they can be used as templates for selective thin film growth, as shown here, allowing lateral confinement of genuine 2D structures, but also can be envisaged as incubators or nanoreactors for specific reactions. So far efforts aiming to tailor crystal surfaces for thin film growth have been mainly directed to substrates with perovskite structure but it is likely that other materials, including rutiles, anatases, and spinels, will probably follow a similar evolution.

**Acknowledgements**

We acknowledge the contribution in some of the works here reviewed of R. Bachelet, N. Dix, L. Garzo´n, G. Herranz, I. C. Infante, M. Paradinas, D. Pesquera and M. Scigaj. Financial support by the Spanish Government (Projects MAT2011-29269-C03, MAT2010-20020 and NANOSELECT CSD2007-00041) and Generalitat de Catalunya (2009 SGR 00376 and 2009 SGR 558) is acknowledged.

# Notes and references

Fig. 1 (a) Sketch of a miscut substrate. L and H are the terrace width and step height, respectively. The corresponding miscut angle a is atan (H/L). Bottom panel: topographic AFM image (3D view) of the surface of an STO single crystal. (b) Top view sketch of a vicinal surface where the steps present an overall in-plane misorientation f with respect to the main crystallographic direction parallel to axis a. Bottom panel: topographic AFM image (2D view) of the same surface shown in (a).

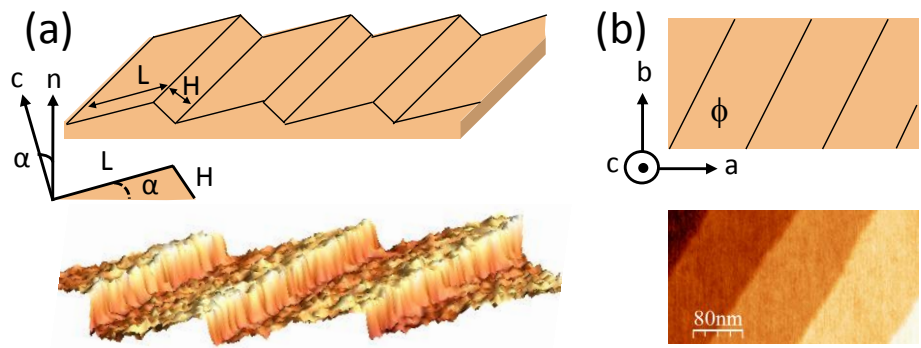

Fig. 2 Sketches of different substrates with (a) intermediate miscut, (b) large miscut or vicinal, (c) very low miscut or singular, (d) step bunching, (e) straight steps (low f), and (f) with kinked steps (high f).

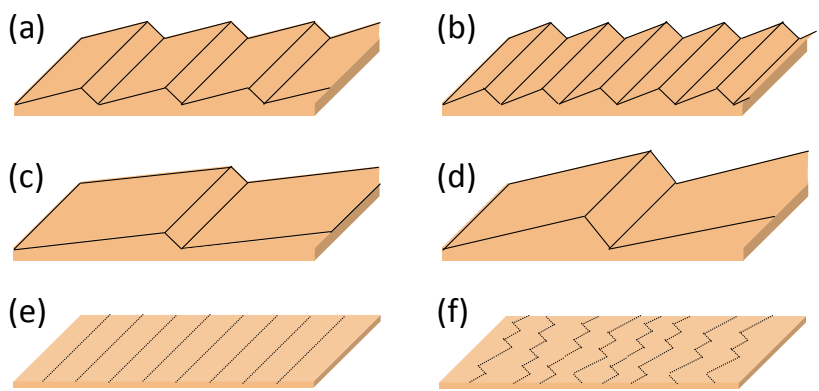

Fig. 3 Sketch of a (001)-oriented perovskite surface exhibiting single chemical termination: AO (a) and $BO_2$ (b). Idem for surfaces in which $BO_2$ and AO (majority and minority terminations, respectively) are randomly distributed (c) or confined along the steps (d).

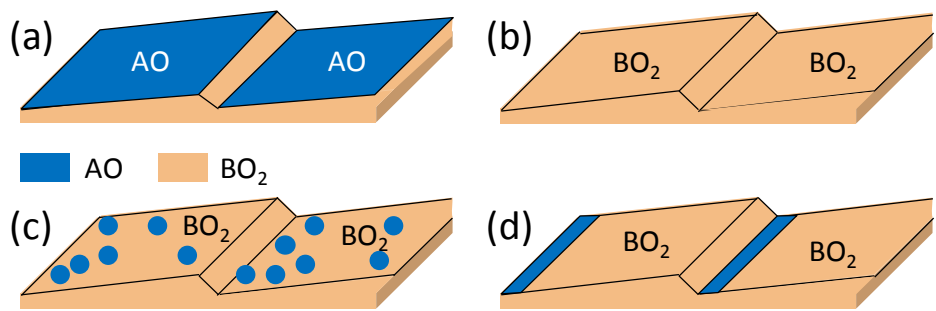

Fig. 4 AFM topographic (left panels) and phase-lag (right panels) images of STO substrates after annealing at 1300 °C in air for (a) 2 h, (b) 12 h, and (c) 72 h. Adapted with permission from Appl. Phys. Lett. (ref. 16). Copyright 2009, American Institute of Physics.

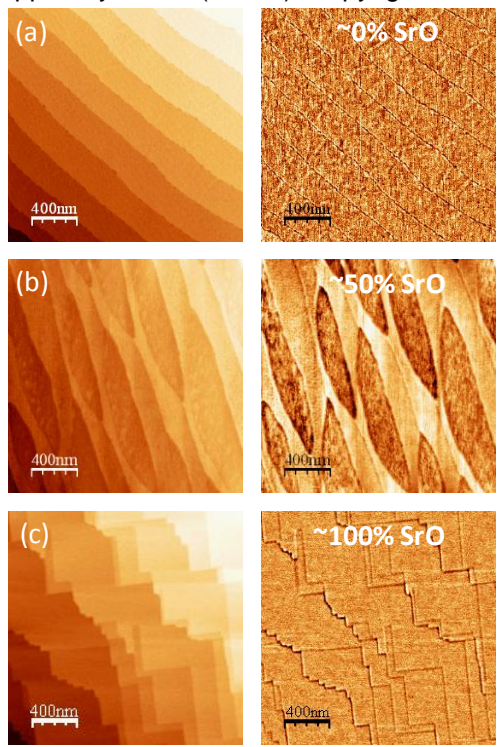

Fig. 5 Top: topographic AFM images for the initial stages of the chemical termination separation and nanostructuring of a STO(001) substrate thermally treated at 1100 1C for 30 min (a, b). The number labeling in (a) and (b) is used to describe the dynamics of the vacancy islands until formation of SrO-terminated patches at the edges of steps separating $TiO_2$ terraces (see text). The height profile along the line in (a) is also shown. Bottom: topographic (left) and phase (right) AFM images after annealing for 2 h. Central panel shows the height profile along the line marked in (c) and a sketch of the deduced surface nanostructure. (a) and (b) are adapted from ref. 18 with permission from the PCCP owner societies; (c) is adapted with permission from Chemistry of Materials (ref. 17). Copyright 2009 American Chemical Society.

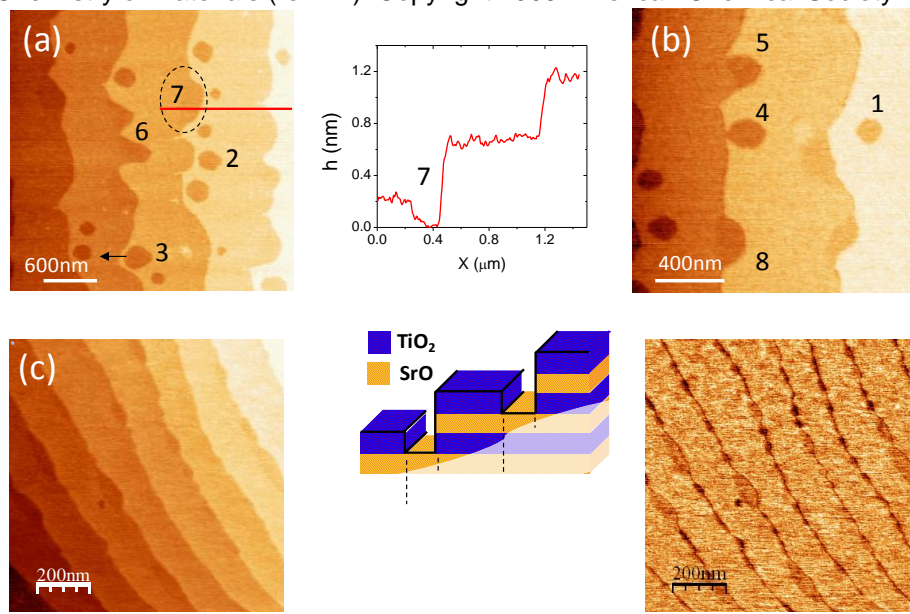

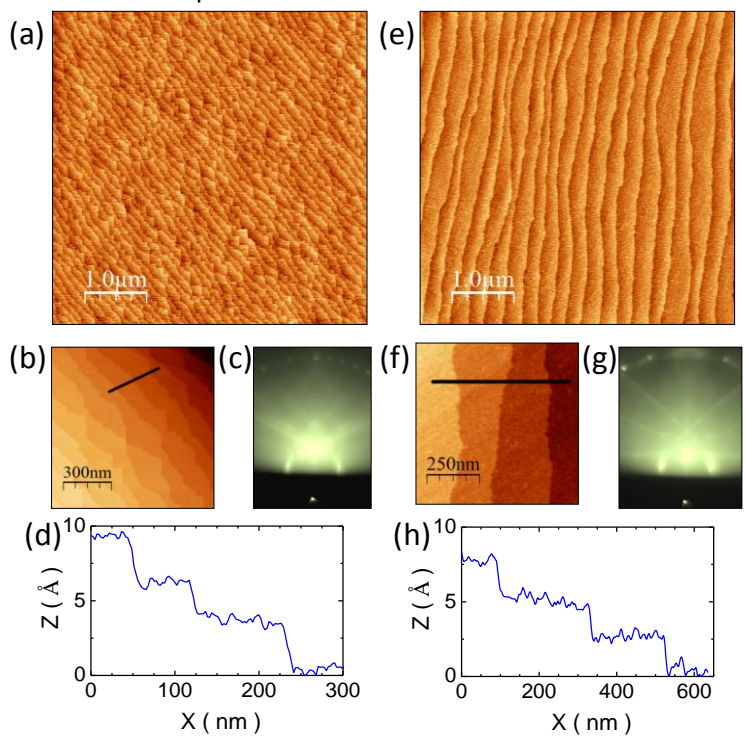

Fig. 6 AFM topographic images of STO(110) (a) and STO(111) (e) substrates treated for 2 h at 1100 °C. Magnified views are shown in (b) and (f). The corresponding height profiles along the marked lines are shown in (d) and (h). RHEED patterns acquired at room temperature and high vacuum taken along [001] and [11-2] are shown in (c) and (g), respectively. The presence of Bragg spots along the 0th and the 1st Laue circles and Kikuchi lines reveal the high quality of the surfaces. Adapted from ref. 21.

Fig. 7 (a) AFM topographic and lateral force (inset) images of a thermally treated LSAT substrate, (b) height profile along the line marked in the AFM image. (d) AFM topographic and phase-lag (inset) images of a 4–5 ML LSMO film deposited on the treated LSAT substrate. (e) Height profile along the marked line (d) and topographic profile along the line in (d). The schematics in the bottom panels illustrate the two chemical terminations (AO, $BO_2$ and A0', $B'O_2$) for each surface: (c) LSAT substrate and (f) LSMO film on LSAT. Adapted with permission from Chemistry of Materials (ref. 32). Copyright 2012 American Chemical Society.

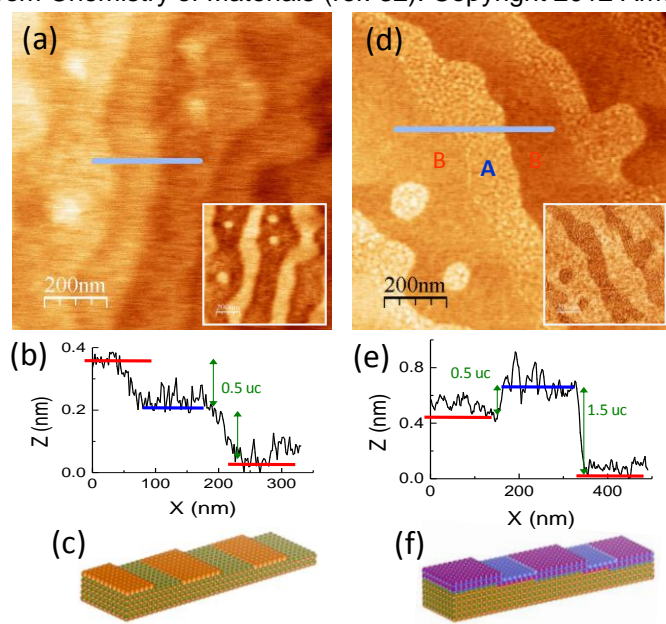

Fig. 8 Topography (a) and the corresponding surface potential (b) of the LSMO surface shown in Fig. 7. AO (A) terminated regions exhibit a higher (lower) surface potential (work function) than regions of $BO_2$ (B) terminated areas. Adapted from ref. 32 with permission from Chemistry of Materials. Copyright 2012 American Chemical Society.

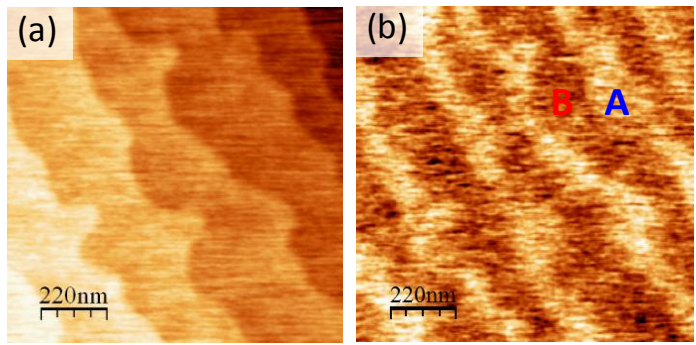

Fig. 9 X-ray linear dichroism spectra of 4 u.c. thick LSMO films grown on $TiO_2$-terminated and SrO-terminated STO(001) substrates. The corresponding AFM topographic images are shown at the panels at the right (top: LSMO on $TiO_2$-terminated STO; bottom: LSMO on SrO-terminated STO). Inset: sketches of the atomic planes in both samples. Adapted from ref. 33.

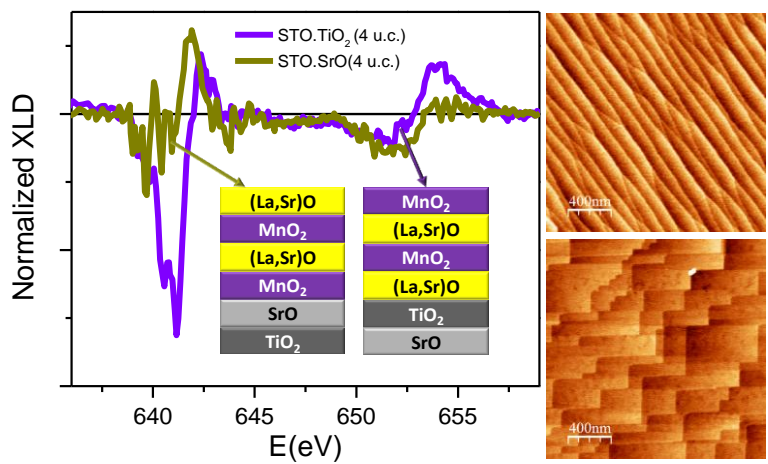

Fig. 10 (a) Scanning electron microscopy image (backscattered electrons) and (b) AFM phase image of a 6 u.c. thick $LaAlO_3$ film grown on a nanostructured STO(001) substrate showing a pattern of $TiO_2$ and SrO terminations. Adapted with permission from Appl. Phys. Lett. (ref. 35). Copyright 2012, American Institute of Physics.

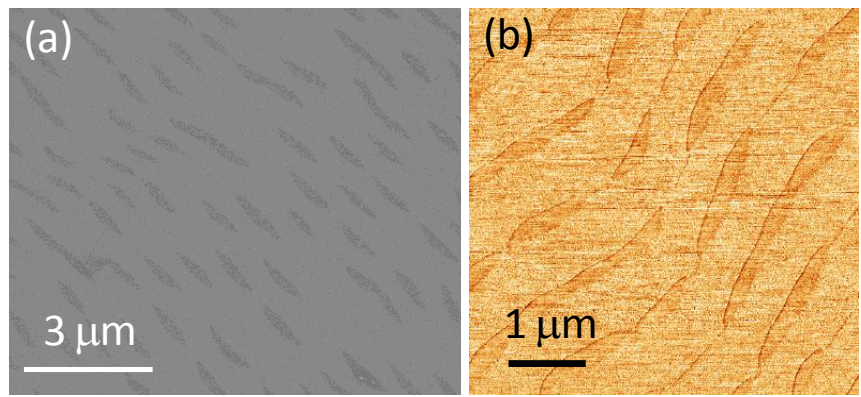

Fig. 11 AFM topographic images of (a) nanostructured STO(001) substrate with minority SrO terminated regions placed along steps. (b) and (c) 1 x 1 mm² and 5 x 5 mm² areas, respectively, of a 7 u.c. thick SrRuO₃ deposited on the surface shown in (a). (d) Electrical conductivity map of the final surface. Inset: sketch of the measuring set-up. Adapted with permission from Chemistry of Materials (ref. 17). Copyright 2009 American Chemical Society.

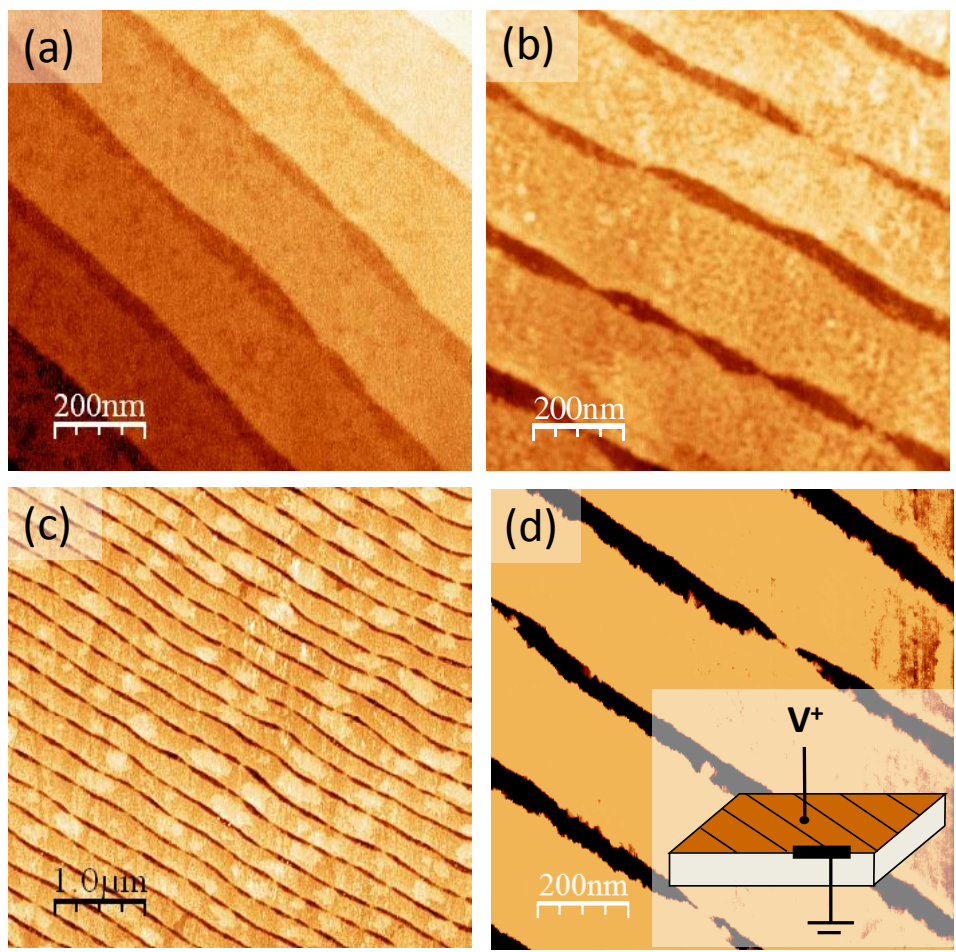

Fig. 12 Topographic AFM image of stearic acid deposited by drop casting onto a nanostructured STO(001) substrate. Right: height profile along the marked line. Adapted from ref. 18 by permission of the PCCP owner societies.

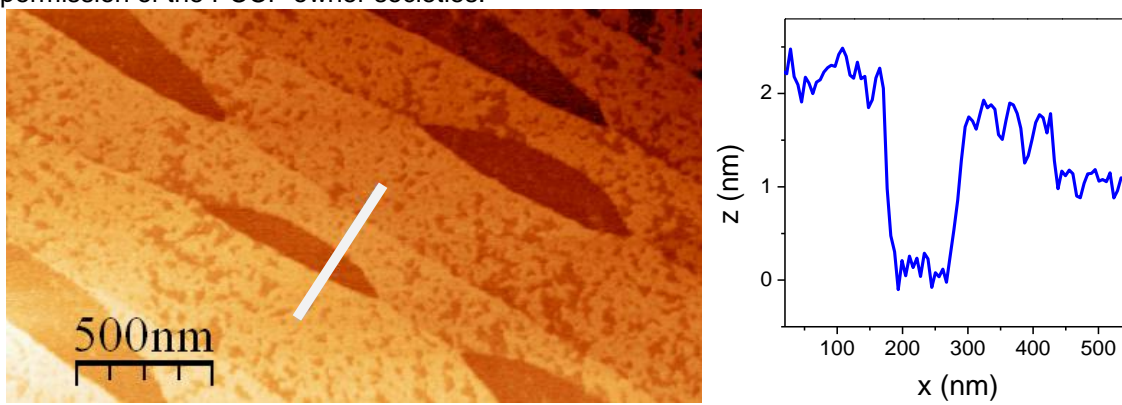

Fig. 13 Strategies to achieve in situ lateral ordering of 0D and 1D nanostructures, including methods of engineered growth (A) and fully self-organized processes (B).

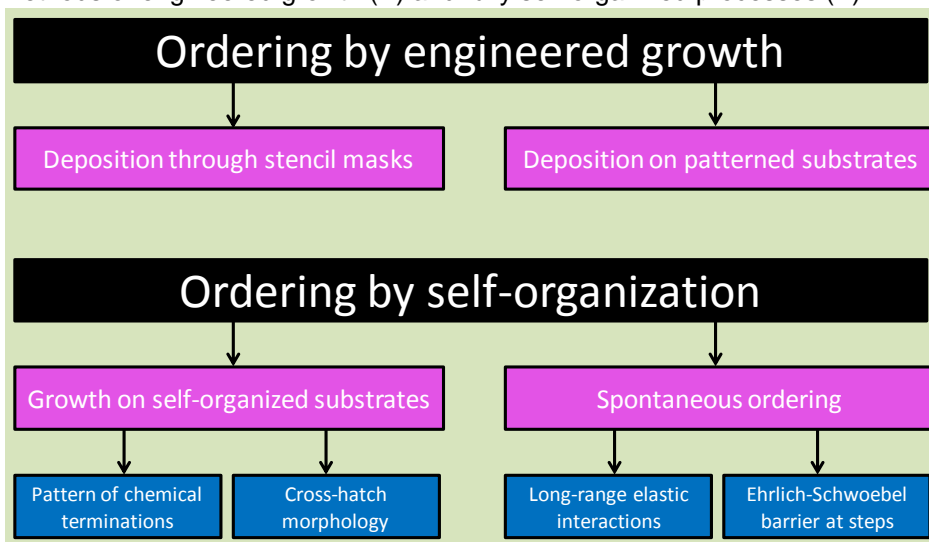

Fig. 14 (a) AFM phase-lag image (1 x 1 mm2) of a nanostructured LSAT(001) substrate. AFM topographic images (b) 1 x 1 mm2 and (c) 5 x 5 mm2 after SrRuO$_3$ growth. The inset in (c) is the self-correlation of the topographic image including the profiles taken perpendicular (inter-line correlation) and parallel (intra-line correlation) to the substrate steps direction. (d and e) Histograms of the lateral sizes of dots shown in (b). (f) Height profile along the line marked in (b). (g) RHEED pattern taken along the [100] direction after SrRuO$_3$ growth. Adapted with permission from Appl. Phys. Lett. (ref. 29). Copyright 2011, American Institute of Physics.

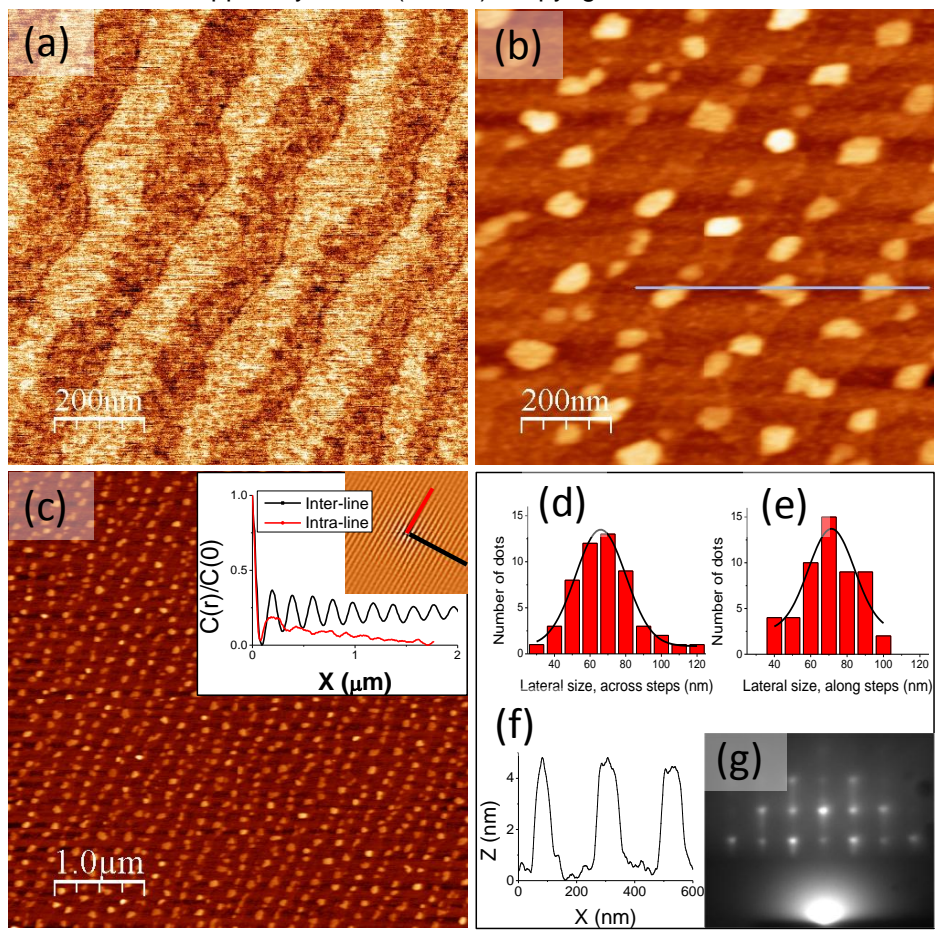